\documentclass[aps,prb]{revtex4}
\usepackage{graphicx}
\usepackage{bm}
\usepackage{xfrac}
\usepackage{amsmath}
\usepackage{amssymb}
\usepackage{ytableau}

\newcommand{\be}{\begin{equation}}
\newcommand{\ee}{\end{equation}}
\newcommand{\bea}{\begin{eqnarray}}
\newcommand{\eea}{\end{eqnarray}}
\newcommand{\beb}{\begin{eqnarray*}}
\newcommand{\eeb}{\end{eqnarray*}}

\begin{document}

\title{Physics of integer spin antiferromagnetic chains~:
Haldane gaps and edge states}

\author{Th.~Jolicoeur}
\author{O.~Golinelli}
\affiliation{Institut de Physique Th\'eorique, Universit\'e Paris-Saclay, CNRS, CEA,\\
 91190 Gif sur Yvette, France}

\begin{abstract}
The antiferromagnetic Heisenberg spin chain with \textit{integer} spin has 
short-range magnetic order and an excitation energy gap above the ground state.
This so-called Haldane gap is proportional to the exchange coupling $J$ of the Heisenberg
chain. We discuss recent results about the spin dependence of the Haldane gap
and conjecture an analytical formula valid asymptotically for large spin values. 
We next study the robustness of the edge states
of the spin one chain by studying by the DMRG algorithm a spin one ladder. We show that
the peculiar hidden topological order of the spin one chain disappears smoothly by increasing the 
ladder rung coupling without any intervening phase transition. This is evidence for the 
fragile character of the topological order of the spin one chain.
\end{abstract}
\maketitle

 

This article is dedicated to Michel Verdaguer, chemist, teacher and colleague \textit{extraordinaire}.
\bigskip

\section{The nonlinear sigma model and the Haldane gap}


It is well understood that the origin of exchange interactions between magnetic
ions in an insulator is essentially of quantum mechanical nature. Indeed it involves
electron delocalization and the Pauli principle as needed for fermionic elementary constituents of matter. 
However the description of magnetic properties of solids at the
mesoscopic or macroscopic scale very often does not involve in a fundamental way
quantum mechanics. One can use coarse-grained magnetization vectors to describe
ordered states, in the ferromagnetic or the antiferromagnetic case that are perfectly classical quantities. 
Also excited states above ordered ground states like magnons admit a classical description which has a wide range of 
applicability. This classical line of thought has been applied for a long time also in the realm of one-dimensional 
quantum spin systems with some success in the case of systems with rather large spin values. So it came as a surprise
when D. Haldane~\cite{haldane_ground_2016,haldane_nonlinear_1983,haldane_continuum_1983} at the beginning of the eighties made the bold 
claim that integer-spin antiferromagnetic chains have a gap to all excitations at odds with previous magnon-based reasonings. 
The picture that emerges is that integer spin chain have only short range magnetic order even at zero temperature with 
spin-spin correlations decaying exponentially with a characteristic length and there is a gap to all excitations. 
On the 
contrary half-integer spin chains have antiferromagnetic order with no characteristic scale albeit decaying as a power law and 
no gap to excited states.
Interestingly the first Haldane gap spin chain NENP~\cite{meyer_crystal_1982} was synthesized roughly at the same time as Haldane's 
conjecture appeared. Since its synthesis NENP has been an extremely good 
platform~\cite{renard_presumption_1987,regnault_wave_1992,renard_quantum_1998} to investigate Haldane gap physics.
Quantitative comparison between theory and experiment has been 
successful~\cite{golinelli_dispersion_1992,golinelli_haldane_1992,golinelli_dynamical_1993}.
Detailed investigations have confirmed the existence of the Haldane gap close to $0.41J$ where $J$ is the nearest-neighor
antiferromagnetic exchange of the isotropic Heisenberg 
chain~\cite{affleck_quantum_1989,golinelli_finite-lattice_1994,white_numerical_1993}.

The Haldane conjecture was originally based on the derivation of an effective quantum field theory using an expansion
of fluctuation about local antiferromagnetic order of a spin-S chain. This derivation has been the subject of much 
theoretical work and is now available in streamlined form in textbooks~\cite{auerbach_interacting_1994}. 
We propose in this section a quick presentation of 
the main arguments and proceed to discuss the dependence of the Haldane gap on the spin magnitude $S$ since there are 
now several numerical studies~\cite{todo_parallel_2018} extending up to $S=4$.

The starting point is the imaginary time representation of the partition function in terms of spin coherent states~:
\be
\mathcal{Z}=\int \mathcal{D}{\hat \Omega}_i
\exp \{iS\sum_i \omega \left[{\hat \Omega}_i \right]-\int_0^\beta S^2
\sum_i {\hat \Omega}_i\cdot {\hat \Omega}_{+1}\}
\label{pathintegral}
\ee
The spin operators $S_i^{x,y,z}$ are written in terms of the spin coherent states
$S_i^{x,y,z}\rightarrow S{\hat \Omega}_i^{x,y,z}$ where now the quantities ${\hat \Omega}_i$
are classical commuting vectors of unit norm and $S$ is the spin magnitude. In fact the Haldane mapping is an 
asymptotic expansion valid for large $S$ and its quantitative used for $S=1$ is not guaranteed. Numerical direct 
studies down to $S=1$ have confirmed its applicability even in this case. The first term 
$\sum_i \omega \left[{\hat \Omega}_i \right]$
in the exponential is the 
\textit{Berry phase} contribution which is a non-trivial function of the imaginary time evolution of the coherent state. 
This term is in fact responsible for the difference between integer and half-integer spin chains. It is irrelevant to the bulk
physics of the integer spin chain case and it is a fortunate circumstance that provided this contribution can be dropped 
then the field theoretical analysis of the effective model we will derive is in fact well-known. From the point of view of 
the effective long-distance low-energy field theory
the spin-1/2 spin chain is much difficult to understand even if this is the only case where
exact results are known for the eigenstates by the Bethe solution. Since we discuss only the Haldane gap we will drop the 
Berry phase form now on (in the integer spin case the Berry phase term is responsible for the appearance of
edge spins S=1/2 as we discuss in the next section).
We now proceed to construct an effective model valid at long distance and low-energies. This requires 
the identification of the modes that are pertinent in this limit. We assume that these are antiferromagnetic as well as 
ferromagnetic fluctuations as observed in all approximation schemes used to study spin chain physics. We write the spin coherent states as~:
\be
{\hat \Omega}_i =(-)^i \, {\hat n}_i \, \sqrt{1-\mathbf{L}^2_i/S^2} 
+\frac{1}{S}\mathbf{L}_i
\ee
where ${\hat n}_i$ is a unit vector and $\mathbf{L}_i$ is a vector perpendicular to
${\hat n}_i$. This way of writing the spin coherent state leads naturally to a $1/S$ expansion. 
The nearest-neighbor Heisenberg coupling is expanded as~:
\be
{\hat \Omega}_i\cdot {\hat \Omega}_{i+1}=
-1 +\frac{1}{2}({\hat n}_i-{\hat n}_{i+1})^2 
+\frac{1}{2S^2}(\mathbf{L}_i+\mathbf{L}_{i+1})^2
\ee
The Berry phase contribution in the path integral leads to a coupling between the
$\mathcal{L}$ vector and the imaginary-time derivative of the order parameter field
$\hat n_i$~:
\be
\int \mathcal{D}{ \mathbf{L}}_i^\perp
\exp \{
-\int_0^\infty d\tau \int \frac{dq}{2\pi}({\hat n}\times\partial_\tau {\hat n})_{-q}\cdot \mathbf{L}_q
-\int_0^\infty d\tau \int \frac{dq}{2\pi}\,\chi^{-1}_{q}\,\mathbf{L}_q\cdot \mathbf{L}_{-q}
\}
\ee
where the susceptibility is given by $\chi^{-1}_{q}=4J\cos^2(qa/2)$ with $a$ the distance between spins. 
Progress can be made if we take the long wavelength 
limit $q\rightarrow 0$ in the susceptibility
$\chi^{-1}\rightarrow 4J\equiv \chi_0$. Then we can integrate out the ferromagnetic fluctuation vector $\mathbf{L}$ and 
obtain a simple quadratic action~:
\be
\mathcal{Z}=\int \mathcal{D}{\hat n}\,
\exp \{-\frac{1}{2}\int_0^\infty d\tau \int dx[
\chi_0 (\partial_\tau{\hat n})^2
+\rho_s(\partial_x{\hat n})^2                                      
]\}
\ee
where we have introduced a stiffness $\rho_s=JS^2a$.
Now Euclidean time and space are on an equal footing provided we rescale by the velocity $c=\sqrt{\rho_s/\chi_0}$. 
At zero temperature $\beta\rightarrow\infty$
this field theory is in fact the well-known nonlinear sigma model. It was introduced
to describe critical properties of classical Heisenberg ferromagnets at nonzero temperature. Here we observe the 
consequence of the mapping of a $1+1$ dimensional quantum system onto a classical 2-dimensional system. In this new 
language there is a very basic fact~: 
there is no order at any nonzero temperature
for classical Heisenberg spins. In physical terms spin-waves disorder the system beyond a some finite correlation 
length $\xi$ and spin correlations decay exponentially with distance. We thus infer immediately that the space 
correlations of the $\hat n$ vector field decay exponentially. Since in addition we have complete equivalence between 
space and (Euclidean) time 
then the time correlations also decay exponentially~:
\be
\langle {\hat n}(0,x){\hat n}(\tau,x)\rangle\approx \exp (-(c\tau)\xi^{-1})
\label{tcorr}
\ee
It is a classic result in Euclidean quantum mechanics that the imaginary-time decay
of correlations is governed by the energy gap of the system. So we conclude from result Eq.(\ref{tcorr}) that 
the spin chain has a gap $\Delta=c\xi^{-1}$. If we rewrite
the effective action in terms of the conventional nonlinear sigma model, we find the action~:
\be
\mathcal{S}_{NL\sigma M}=\frac{1}{2g}\int d\tau dx\, (\partial_\mu {\hat n})^2
\ee
with the coupling constant $g=c/\rho_s=2/S$.
Renormalization group calculations have shown that correlation length of this model
has a known dependence upon $g$~:
\be
\xi = C\, \frac{2\pi}{g}\,  \exp (-2\pi/g)(1+O(g)),
\ee
where $C$ is a pure number. Since we know the spin dependence of the velocity
$c=2JSa$ we deduce that the gap depends upon the spin value as~:
\be
\Delta_S/J = C_0 S^2 \exp (-\pi S)(1+O(1/S)),
\ee
with a prefactor $C_0$ which is a pure number independent of the spin. This formula should be understood as being asymptotic i.e. 
valid when $S\rightarrow\infty$.
The case S=2 has been studied experimentally~\cite{granroth_experimental_1996} and theoretically~\cite{schollwoeck_s2_1996}.
More recent theoretical estimates are available~\cite{todo_cluster_2001} for S=3 and even~\cite{todo_parallel_2018} for S=4.

\begin{table}[ht]
\begin{center}
\resizebox{10cm}{!}{
 \begin{tabular}{|r|c|c|c|c|c}\hline
  S  &  Gap/J &    NL$\sigma$M &   Gap/NL$\sigma$M & Shanks \\ \hline
  1 &  4.104800e-01 &  4.321392e-02 &   9.499  &  --    \\
  2 &  8.916000e-02 &  7.469771e-03 &  11.936  &  --    \\
  3 &  1.002000e-02 &  7.262957e-04 &  13.796  & 19.787 \\
  4 &  7.990000e-04 &  5.579748e-05 &  14.320  & 14.525 \\ \hline 
  \end{tabular}
}
\end{center}
\caption{Gaps of the spin-$S$ Heisenberg chain using state-of-the-art numerical results~\cite{todo_parallel_2018}. 
The first column gives the spin value, the second column the most accurate known values 
of the dimensionless gap, the third column the bare $\sigma$ model number from which we compute ratio in the fourth column. It should 
extrapolate to a constant for infinite spin. The extrapolation
is performed in the last column using the simple three-term Shanks extrapolation.}
\label{gaps-gaps}
\end{table}

The pure number in factor of the $\sigma$ model formula for $\xi$ is
known and is $C=e/8$ for a specific choice of the renormalization procedure.
The renormalization procedure corresponding to the quantum lattice model cannot be easily related to the standard ways of renormalizing 
the $\sigma$ model so since we are not yet able to make this correspondence precise we choose to make an extrapolation from known values
of the Haldane gap up to $S=4$~: see Table I.
The Shanks extrapolation for three numbers $a,b,c$ is simply given by the ratio
$(ac-b^2) / (a+c-2b)$ and is given in the last column of table I. After some trial and error we are led to propose that the limiting value is 
  $2\exp(2) = 14.778...$ close to 14.525. We thus conclude this section by conjecturing the asymptotic formula for the Haldane gap~:
 \be
 \Delta_S/J=2 S^2 \exp (2-\pi S)(1+O(1/S))
 \ee
It remains to be seen if methods like those of ref.\cite{jolicoeur_about_1991} can relate the renormalization schemes and 
confirm this prefactor.

\section{Edge states of a Haldane ladder}

Another important feature of the spin-1 chain is the existence of a hidden magnetic order which is long-range. We have already 
commented on the fact that ordinary spin correlations decay exponentially with a finite correlation length. However there is 
nevertheless a hidden order in the system. The nearest-neighbor Heisenberg S=1 spin chain is not a solvable system and hence 
one has to use numerical methods to study its quantitative properties. However it is possible to perturb the Hamiltonian and 
obtain a new model~\cite{affleck_rigorous_1987} called AKLT whose ground state wavefunction can be found by elementary means even though it 
is still not solvable. The 
trick is to add a biquadratic spin coupling~:
\be
\mathcal{H}_{AKLT}=\sum_i {S_i}\cdot {S_{i+1}}+\frac{1}{3}({S_i}\cdot {S_{i+1}})^2 .
\label{AKLT}
\ee
The motivation is formal since in the real world superexchange theory leads to negligible higher-spin couplings. 
However with the very special choice of the 1/3 coefficient for the biquadratic term the spin operator is now the projector onto
total spin $S=2$ of the pair $i,i+1$. This can be checked by elementary means (a bit tedious). Now we proceed to 
exhibit a wavefunction which is a exact zero-energy eigenstate of Hamiltonian (\ref{AKLT}). Each local spin-1 can 
be viewed as the triplet state of two spin-1/2 residing at the same site $i$. In the case of NENP this corresponds 
to microscopic physics since spin-1/2 elementary electrons in the Nickel ions are coupled by Hund's to a triplet state. 
Now we make singlet bonds between neighboring sites (ions) and construct the wavefunction by capturing all spins in this way~:
\be
|\Psi_{AKLT}\rangle=\dots\otimes \{|\uparrow_i\downarrow_{i+1}\rangle
-|\downarrow_i\uparrow_{i+1}\rangle\} \otimes\{
|\uparrow_{i+1}\downarrow_{i+2}\rangle
-|\downarrow_{i+1}\uparrow_{i+2}\rangle\}\otimes
\dots
\ee
If we consider two neighboring spins then the total spin of such a pair can only be 0 or 1 but not 2 because 
the central spins 1/2 are already engaged in a singlet state. So we see that all projectors onto the S=2 state 
for any pair gives simply zero and hence this wavefunction is an exact zero-energy eigenstate of the Hamiltonian Eq.(\ref{AKLT}).
In fact this is the exact ground state of the AKLT Hamiltonian.
While this construction may seem artificial we note that this Hamiltonian can be viewed as a perturbation of the usual 
Heisenberg Hamiltonian provided ``1/3 is small'' meaning that the biquadratic operator does not change the physics 
of the system. Indeed we are lucky and it has been shown in detail that this is the case~\cite{schollwoeck_onset_1996}. 
As a consequence reasonings 
based on the model AKLT state are likely to be correct for the real-world dominated by Heisenberg exchange. A very simple 
inference from the AKLT description is that if we consider an open chain then there are dangling spins 1/2 at the end. 
Of course real materials always involve open chains and we thus can predict that local probes should detect effective 
spin 1/2 degrees of freedom at the end of NENP chains. The experiment has been done~\cite{hagiwara_observation_1990} 
using NMR of a Zinc impurity breaking the chain and the number of NMR lines is exactly in agreement with the prediction 
of the spin 1/2 instead of 1 as one may guess naively. For an open chain we expect that the two end spins 1/2 appear as
almost degenerate singlet and triplet states due to a coupling through the bulk of the chain that should go exponentially
to zero when the chain length increases. So the spectrum of an open chain should display quasidegenerate S=0 and S=1
states and, above a (Haldane) gap, a S=2 excitation obtained by creating a magnon-like state~\cite{kennedy_exact_1990}. 
If we look at the magnetization profile
of the $S^z=1$ member of the triplet it should display non-zero magnetization only close to the edges of the chain.
This is indeed what is observed numerically for the S=1 chain using the 
DMRG algorithm~\cite{white_numerical_1993,schollwoeck_density-matrix_2011}.


\begin{figure}
\centering
\includegraphics[width=0.6\columnwidth]{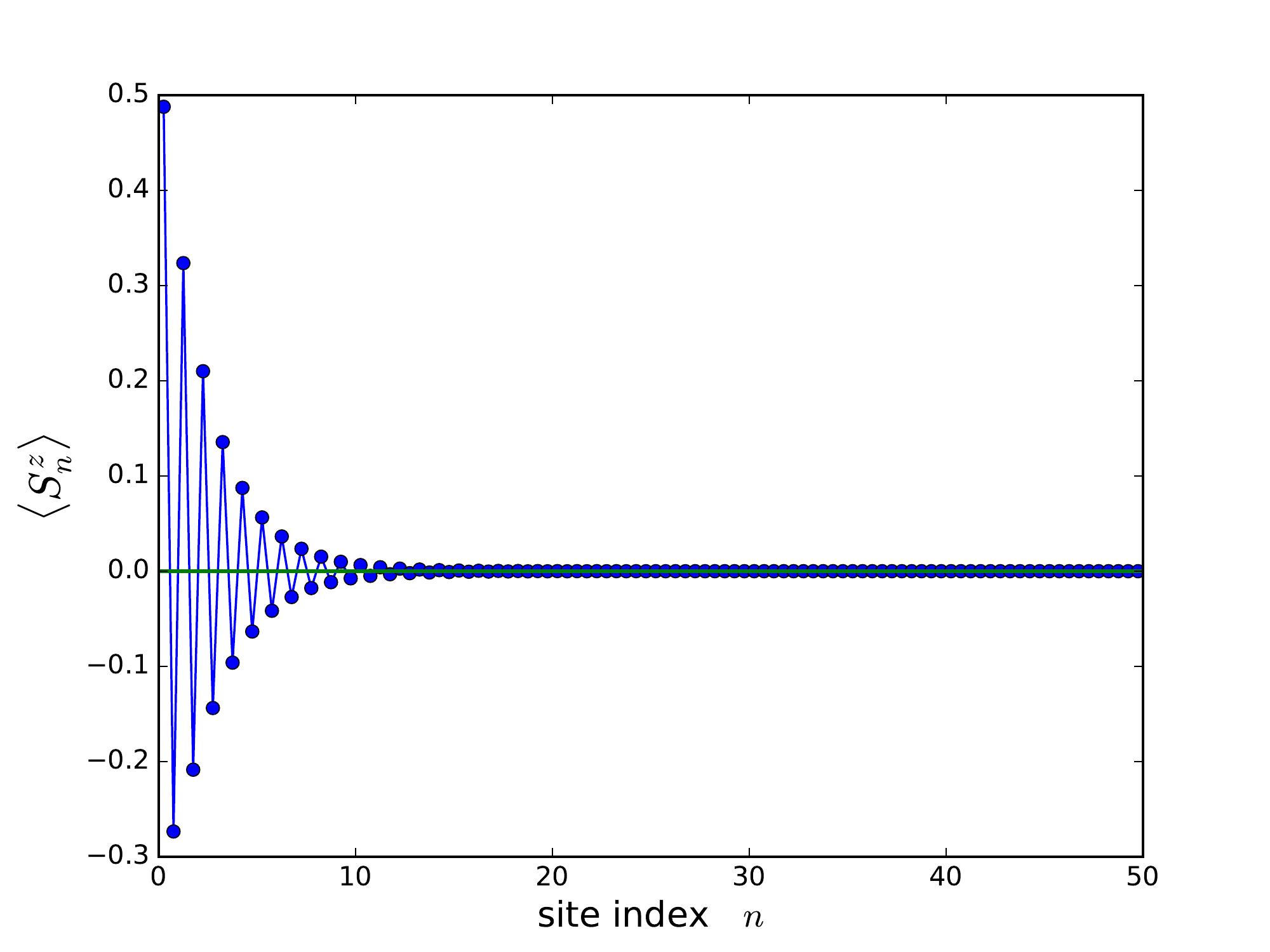}
\caption{The magnetization profile $\langle S^z_n \rangle$ for site $n$ along the first 50 spins of a 
S=1 Heisenberg ladder of 100 sites long (for a total of 200 spins).
The expectation value is computed in the lowest-lying $S^z=1$ state which belongs to the manifold of edge states.
The rung exchange coupling is $J_\perp=0.003$. The green line is zero magnetization. The bulk is not magnetized
proving the edge nature of the excitations.}
\label{lad1-003}
\end{figure}

\begin{figure}
\centering
\includegraphics[width=0.6\columnwidth]{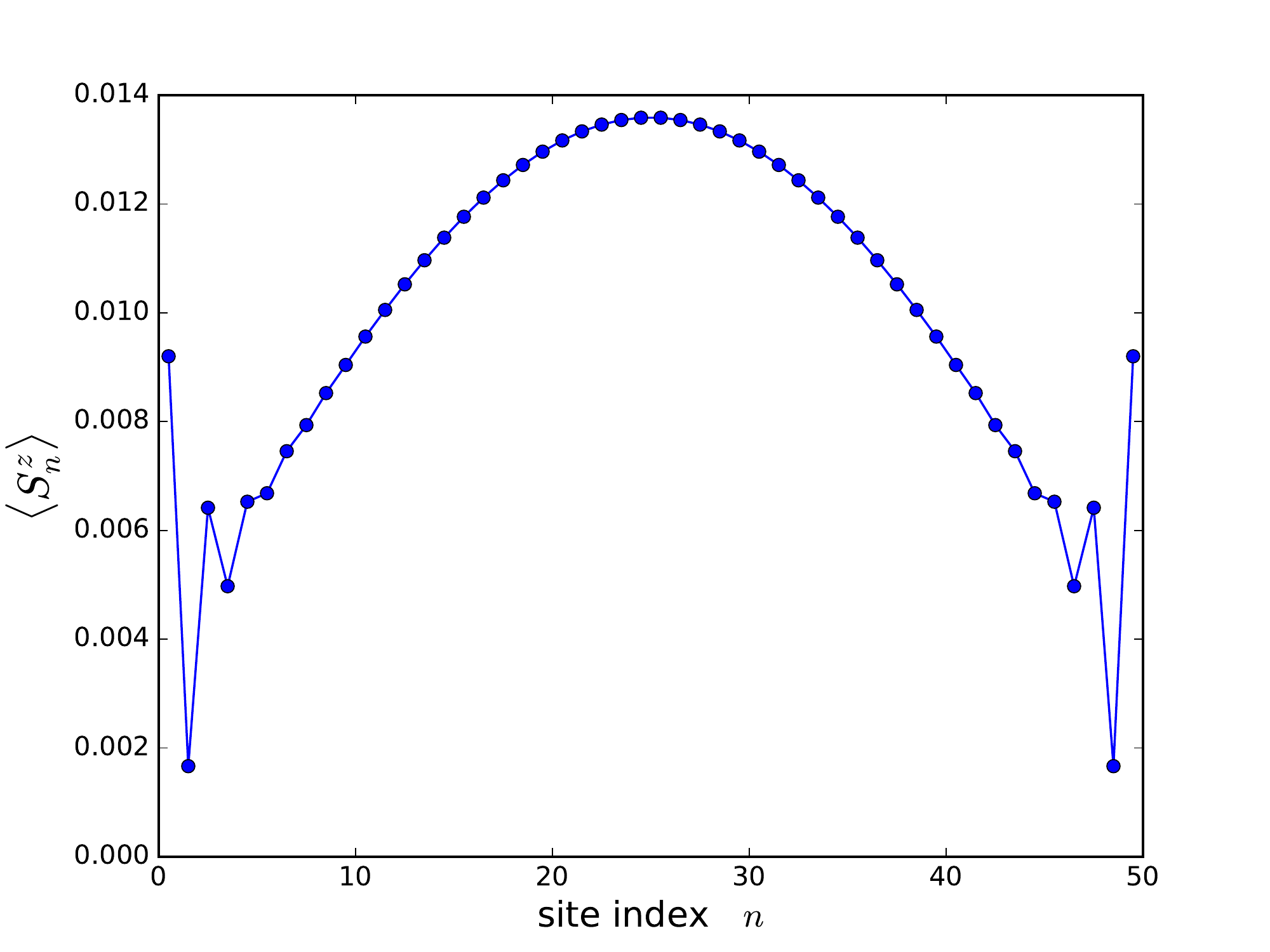}
\caption{The magnetization profile $\langle S^z_n \rangle$ in the lowest-lying $S^z=1$ state for site $n$
along the S=1 Heisenberg ladder of 50 sites long (total 100).
The rung exchange coupling is now $J_\perp=0.5$~: we observe that the magnetization is now essentially in 
the bulk and there are still
edge oscillations of the magnetization.}
\label{lad1-05}
\end{figure}

\begin{figure}
\centering
\includegraphics[width=0.6\columnwidth]{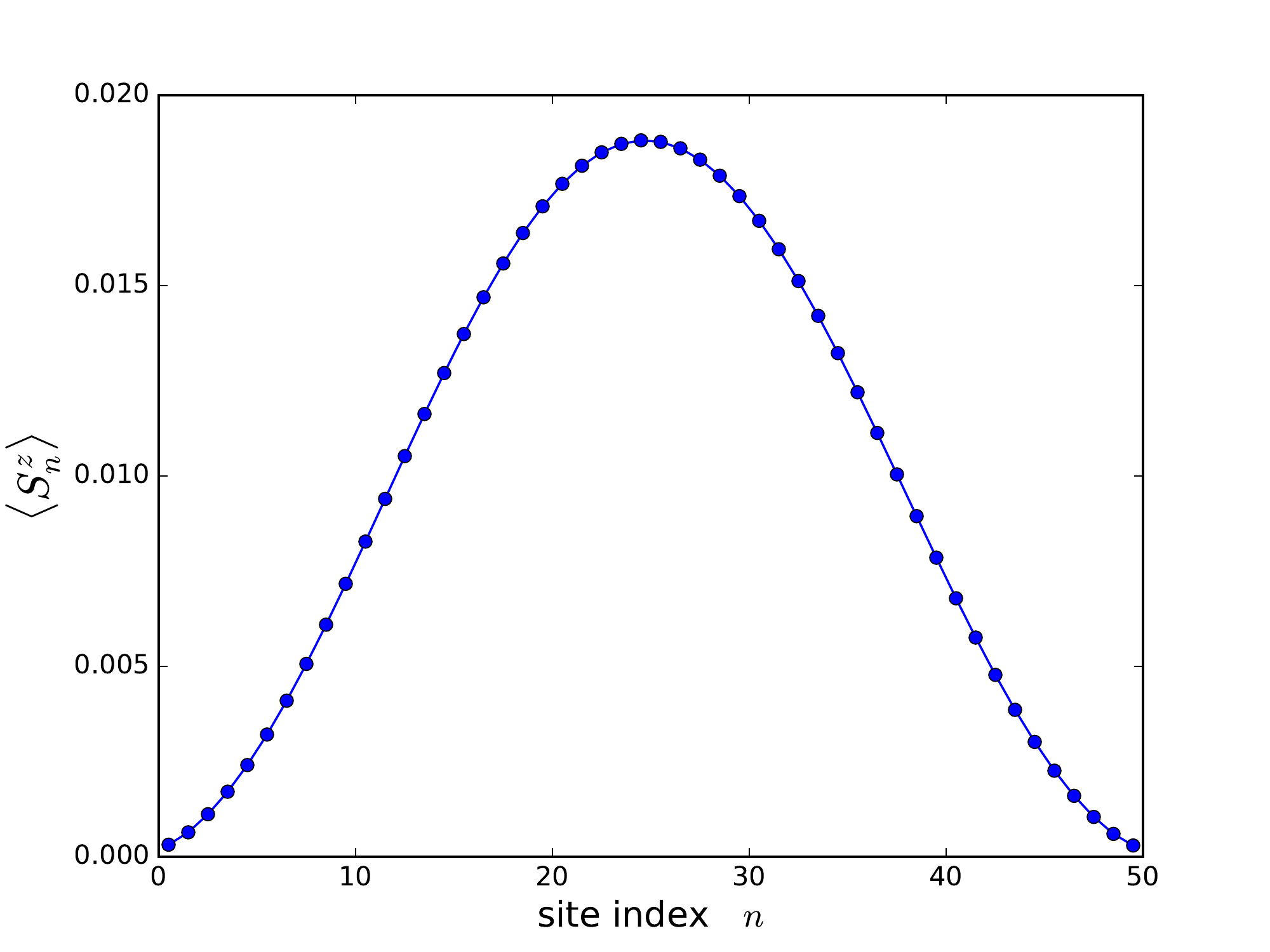}
\caption{The magnetization profile as in fig.(\ref{lad1-05}).
The rung exchange coupling is $J_\perp=4$~: the state can be interpreted as a single magnon state clearly 
defined from the infinite rung limit.
There is no reminder of the edge modes.}
\label{lad1-4}
\end{figure}

We now turn to the question of the robustness
of the formation of these end spin states. They are due to the special ordering pattern described by the AKLT 
wavefunction~\cite{affleck_rigorous_1987,den_nijs_preroughening_1989}
and it is still present for the realistic Heisenberg exchange Hamiltonian.
It is known that several physical effects can destroy the Haldane gap like too strong anisotropies either local on-site or exchange
and also an applied magnetic field can close the Haldane gap. If we cross a phase transition by such a mechanism we do not expect
edge states to survive since the peculiar long-range correlations of the AKLT state are destroyed. It has been noted 
some time ago that in fact there is no need of a sharp phase transition to kill the hidden order and the edge states.
This is very clear in the case of the magnetization process of the Heisenberg S=1 chain. Indeed under an applied field
the Haldane gap goes to zero at some critical field $H_c$ and the magnetization remains zero up to this value.
When the critical value is exceeded the magnetization starts to rise continuously up to full saturation and the edge
states disappear right at $H_c$. If now we consider a realistic material like NENP then there are small but nonzero
anisotropies that break the spin rotation symmetry. As a consequence the magnetization transition~\cite{katsumata_magnetization_1989} 
is now rounded and is no
longer sharp. Magnetization appears immediately for infinitesimal applied fields and edge states disappear when going to
saturation without any phase transition. This phenomenon has been dubbed~\cite{fuji_effective_2016} ``symmetry protected topological order''.
It leaves open the possibility that the hidden order and accompanying edge states may disappear even without breaking any symmetry
and without any phase transition.  We now show that this is the case by studying a coupled S=1 ladder system.
We thus envision a plausible molecular magnet in which neighboring Haldane chains of spins 
S=1 are coupled along the rungs
of a ladder by some independent exchange $J_\perp$. A spin Hamiltonian can thus be written as follows~:
\be
\mathcal{H}_{ladder}=\sum_i {S_{i,A}}\cdot {S_{i+1,A}}+\sum_i {S_{i,B}}\cdot {S_{i+1,B}}
+J_\perp \sum_i {S_{i,A}}\cdot {S_{i,B}},
\label{Hladder}
\ee
where the sites along the two chains are labeled by $i$ and $A,B$ label the two coupled chains. We have taken
the exchange along the chain as the unit of energies so the only remaining parameter is the ratio of exchange
interactions along and across the chains called $J_\perp$. 
We note that recent experiments have shown that the organic molecular magnet BIP-TENO is precisely 
a spin-1 ladder~\cite{katoh_singlet_2000,sakai_magnetization_2003}. However the exchange interactions in BIP-TENO are more complex
than our simple Hamiltonian Eq.(\ref{Hladder}).

If we consider a ladder with periodic boundary conditions  then it has been established that there is no phase 
transition~\cite{allen_spin-1_2000,todo_plaquette-singlet_2001}
between the $J_\perp=0$ decoupled chain limit and the $J_\perp=\infty$ with extremely strong rungs.
This strong rung limit is simple to analyze~: the ground state is made of singlet states involving two spins that are
related by the rung coupling and the total state is just the tensor product of such singlets. The first excited state
is made by breaking a rung bond from singlet to triplet and this triplet will have a dispersion along the chain
whose magnitude is given by the (relatively) small coupling along the chains. The small $J_\perp$ limit is two weakly
coupled Haldane chains and since they are gapped they will be resilient to any small local perturbation like a small
rung coupling. Even if there is no transition between these two limits it remains unclear what happens to the edge states
of an open ladder.

\begin{figure}
\centering
\includegraphics[width=0.6\columnwidth]{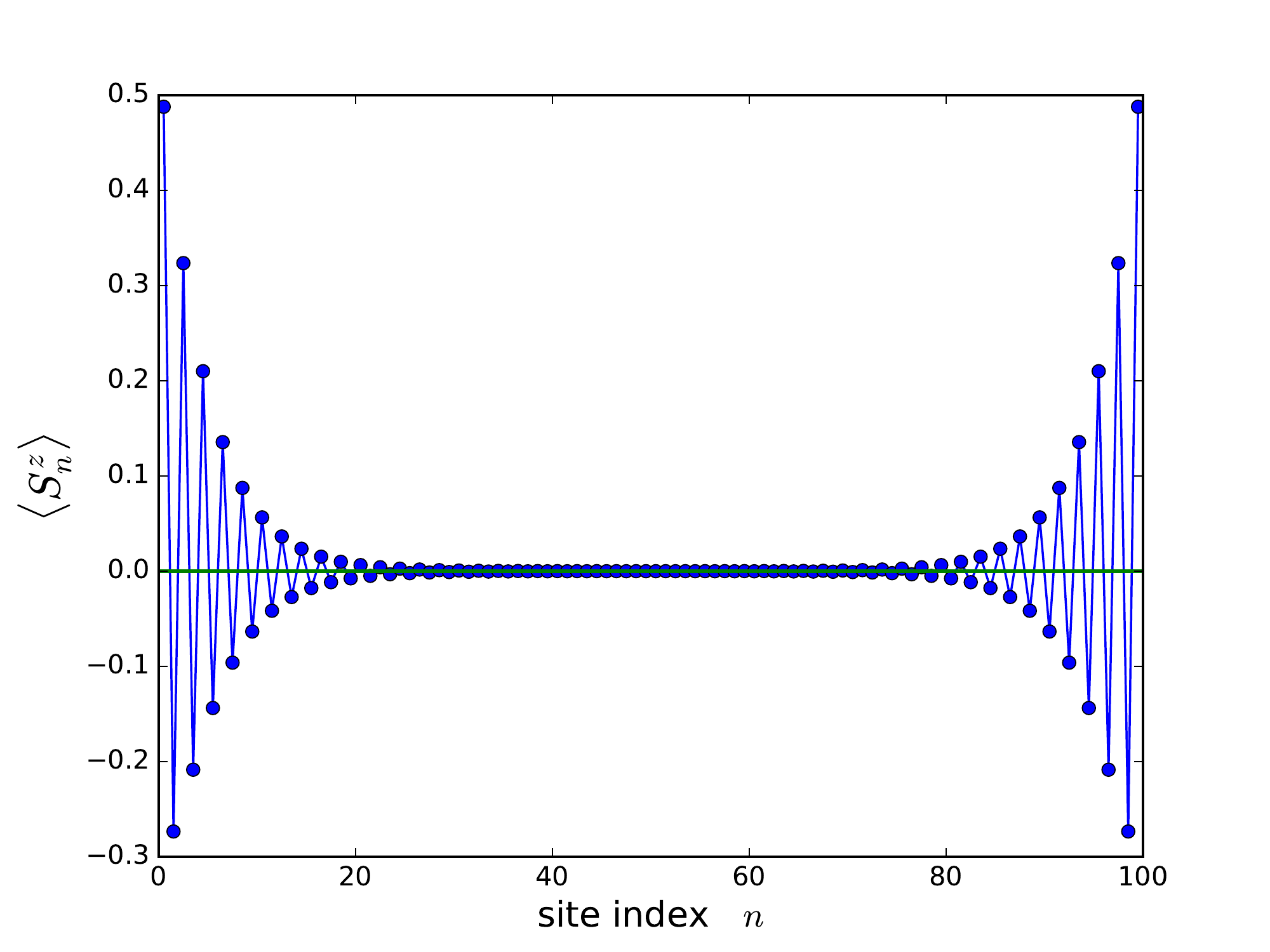}
\caption{The magnetization profile $\langle S^z_n \rangle$ for site $n$ along the S=1 Heisenberg ladder of 100 sites long.
The expectation value is computed in the lowest-lying $S^z=2$ state which belongs to the manifold of edge states.
The rung exchange coupling is $J_\perp=0.003$~: there is zero bulk magnetization and only edge modes.}
\label{lad2-003}
\end{figure}

\begin{figure}
\centering
\includegraphics[width=0.6\columnwidth]{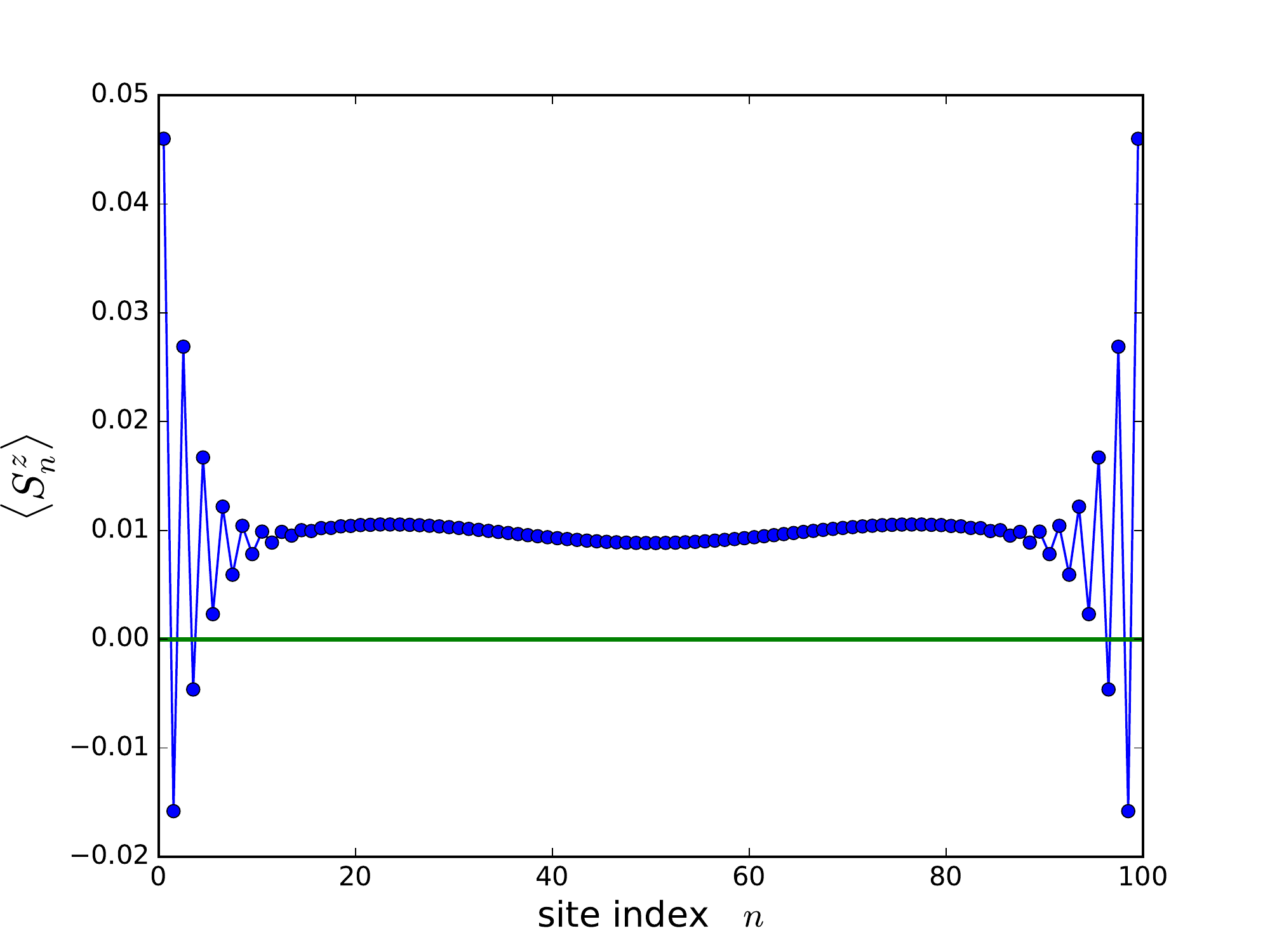}
\caption{The magnetization profile $\langle S^z_n \rangle$ as in fig.(\ref{lad2-003}).
The rung exchange coupling is $J_\perp=0.015$~: there is now coexistence of a small bulk magnetization and edge modes.}
\label{lad2-015}
\end{figure}

\begin{figure}
\centering
\includegraphics[width=0.6\columnwidth]{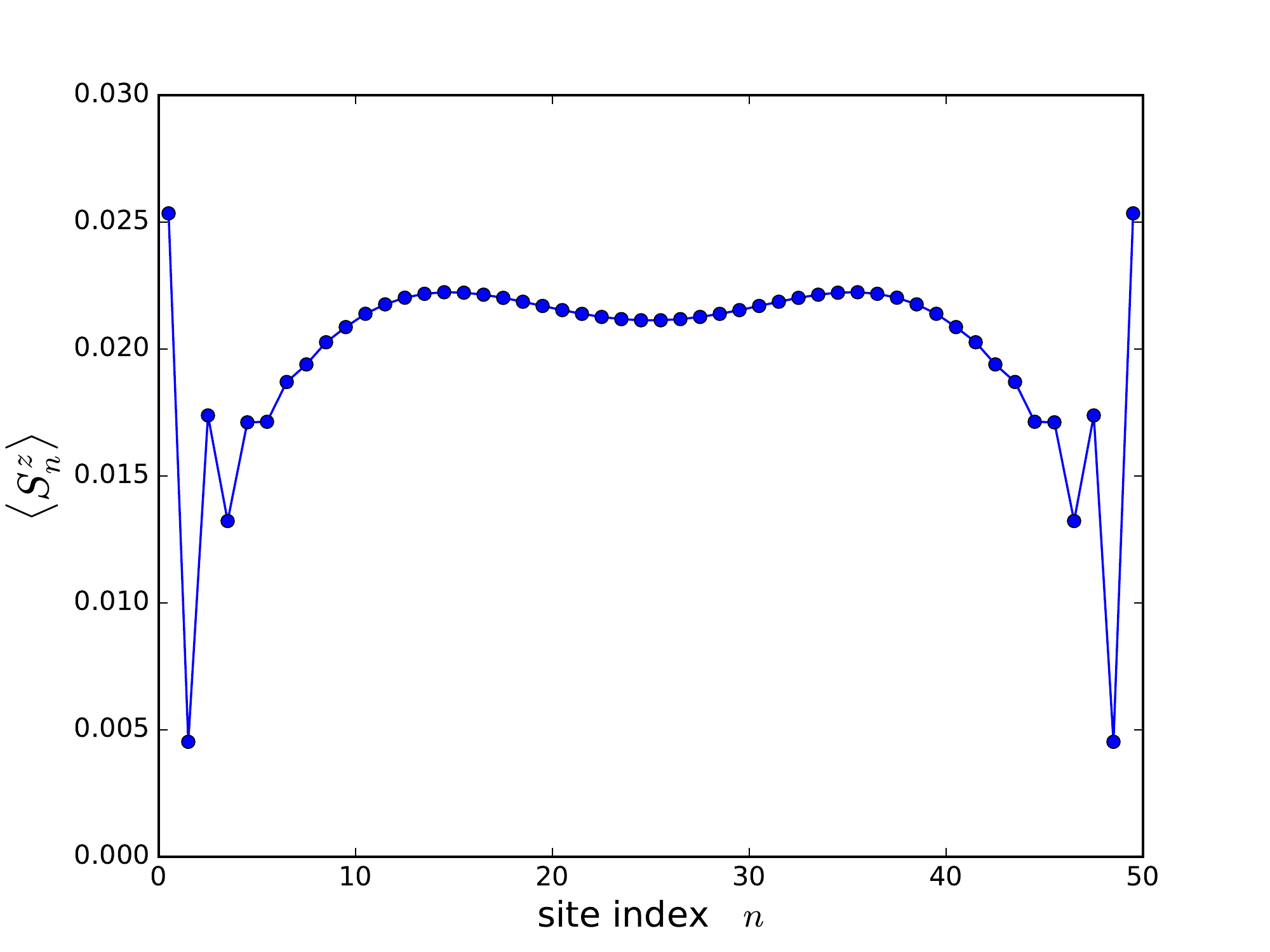}
\caption{The magnetization profile $\langle S^z_n \rangle$ as in fig.(\ref{lad2-015}).
The rung exchange coupling is $J_\perp=0.5$~: the bulk excitations display a two rounded peak structure.}
\label{lad2-05}
\end{figure}

\begin{figure}
\centering
\includegraphics[width=0.6\columnwidth]{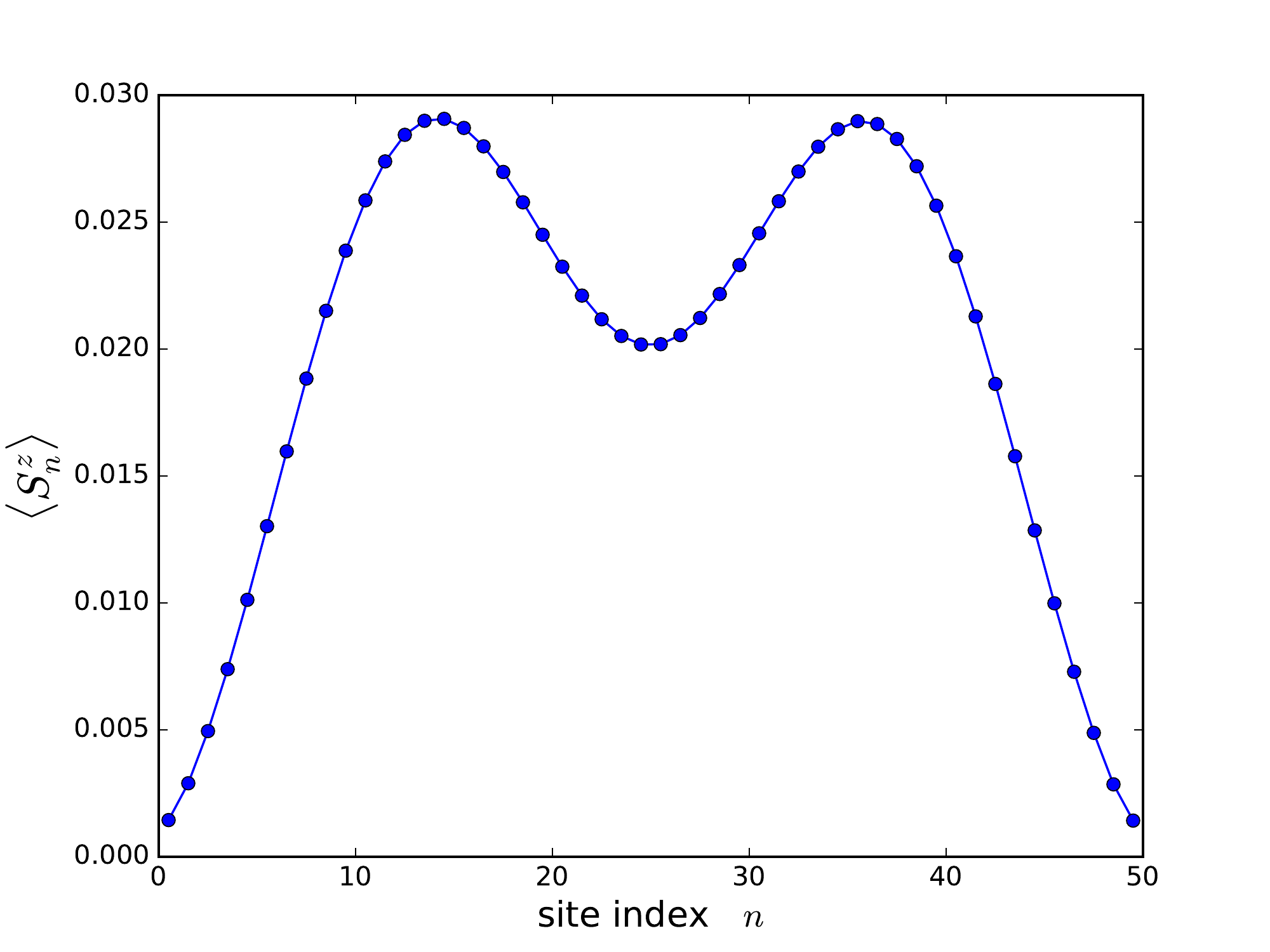}
\caption{The magnetization profile $\langle S^z_n \rangle$ as in fig.(\ref{lad2-05}).
The rung exchange coupling is $J_\perp=4$~: edge modes have disappeared completely and we observe two magnons
with characteristics of a particle-in-a-box wavefunction.}
\label{lad2-4}
\end{figure}

\begin{figure}
\centering
\includegraphics[width=0.6\columnwidth]{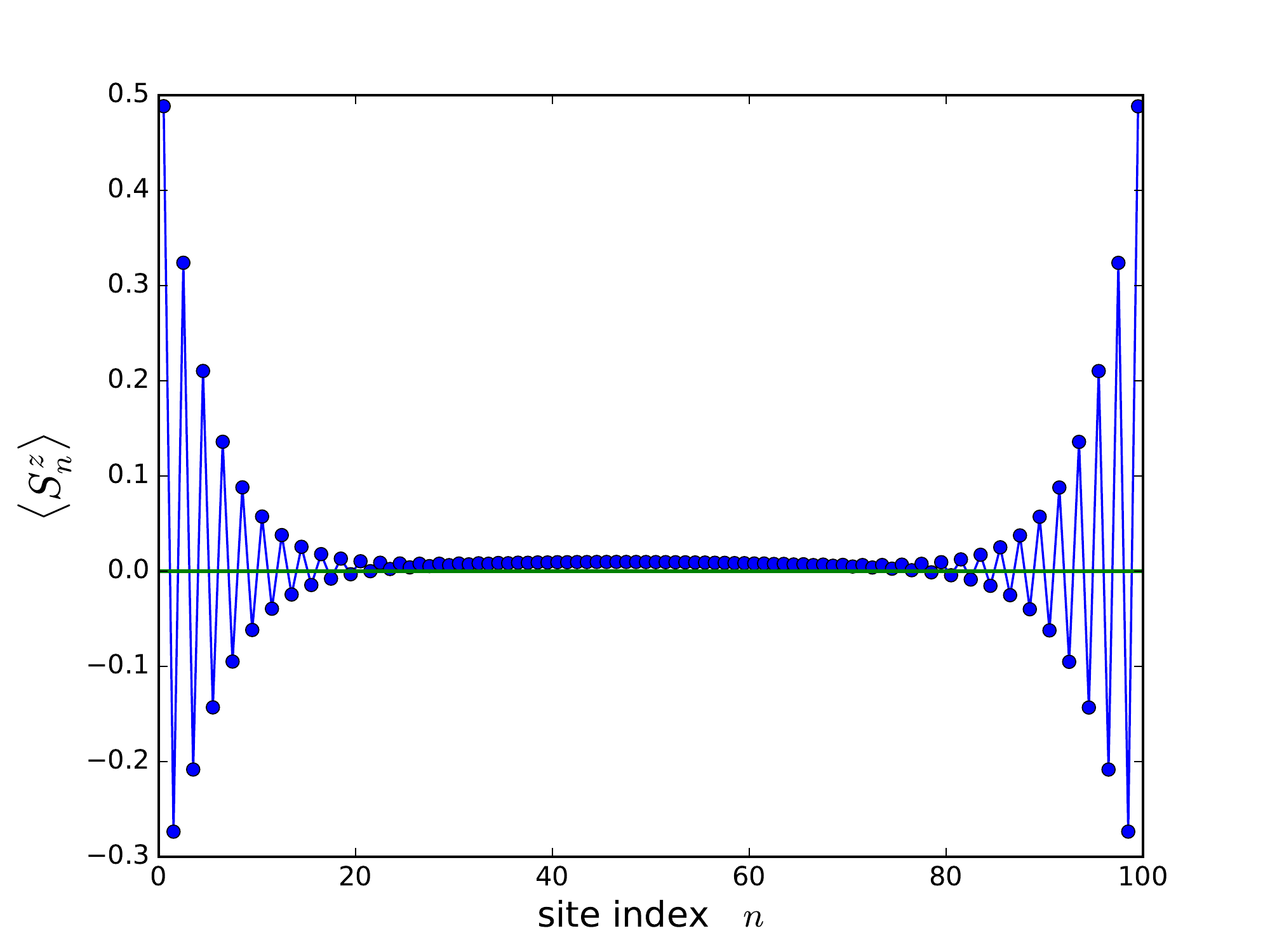}
\caption{The magnetization profile $\langle S^z_n \rangle$ in the lowest-lying $S^z=3$ state
The rung exchange coupling is $J_\perp=0.003$~: while there are edge oscillations we note that the bulk magnetization
is nonzero. This is not a pure edge state.}
\label{lad3-003}
\end{figure}

\begin{figure}
\centering
\includegraphics[width=0.6\columnwidth]{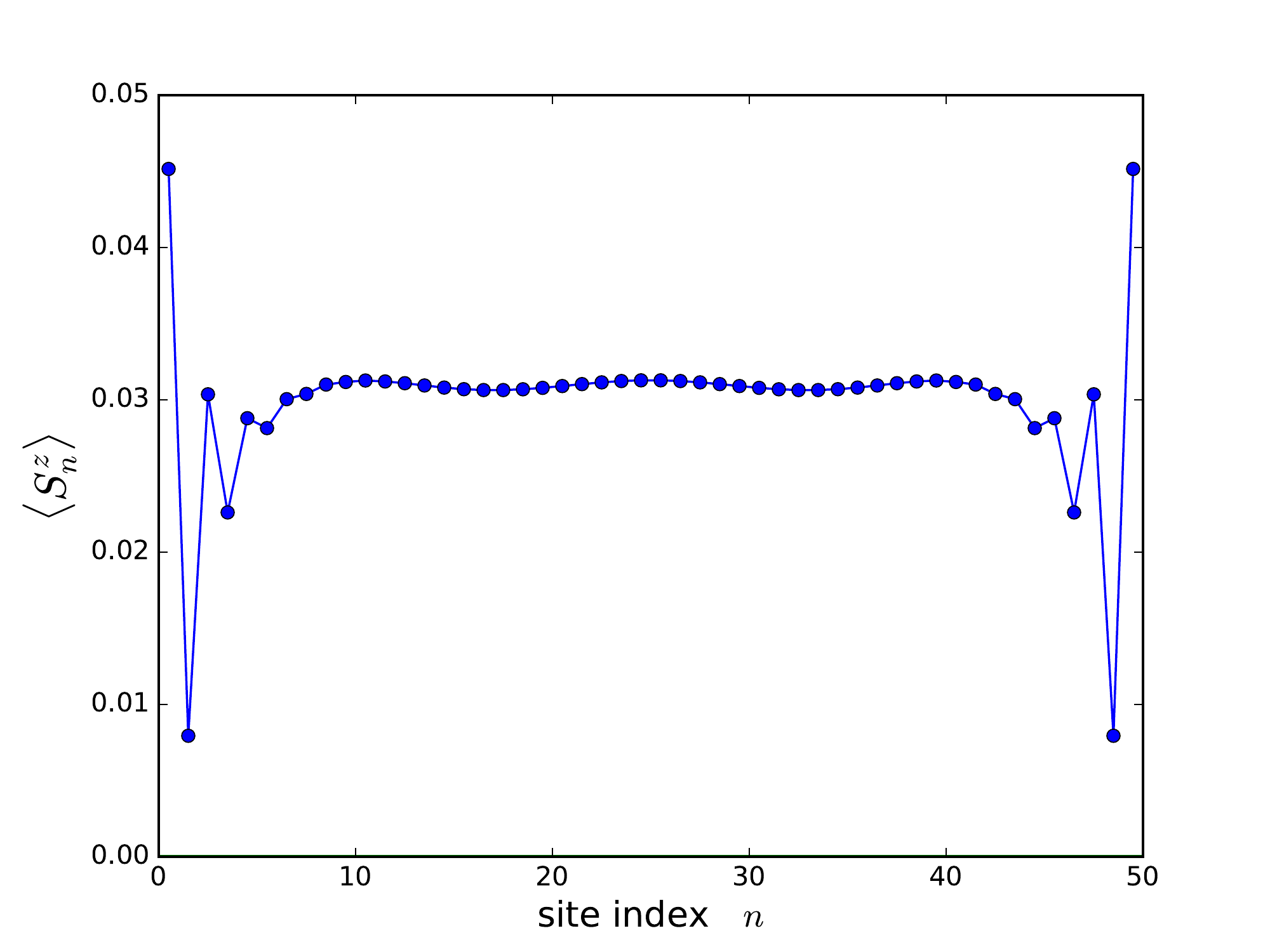}
\caption{The magnetization profile $\langle S^z_n \rangle$ in the lowest-lying $S^z=3$ state .
The rung exchange coupling is $J_\perp=0.5$~: edge oscillations are still present but the three-peak structure appears.}
\label{lad3-05}
\end{figure}

\begin{figure}
\centering
\includegraphics[width=0.6\columnwidth]{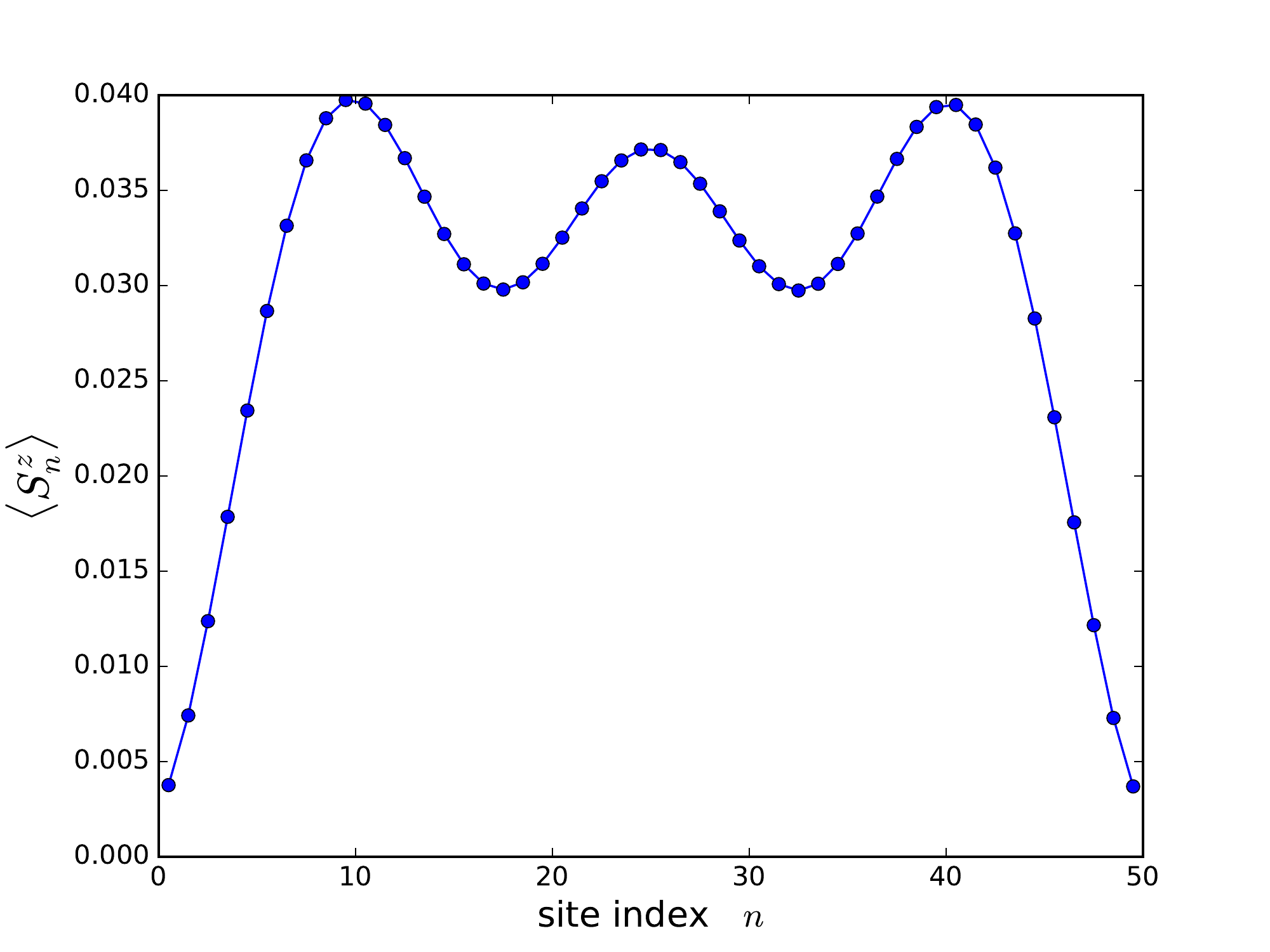}
\caption{The magnetization profile $\langle S^z_n \rangle$ in the lowest-lying $S^z=3$ state.
The rung exchange coupling is $J_\perp=4$~: we observe a pure  three-magnon state.}
\label{lad3-4}
\end{figure}

We first enumerate the low-lying states of this spin system
starting from our knowledge of the S=1 chain. We known that each chain will have a singlet and a triplet of low-lying states
separated by the Haldane gap from the higher excitations~\cite{kennedy_exact_1990}. 
With two chains this means a total of 16 states that can be classified
as two S=0 singlets, three S=1 triplets and one S=2 quintuplet.
When adding a very small $J_\perp$ these will no longer be degenerate and first-order perturbation theory will lead
to an order $O(J_\perp)$ splitting of the states. Their edge nature can be revealed by computing the magnetization profile
$\langle S^z_n\rangle$ along one of the coupled chains. We have performed DMRG calculations of this system to obtain
the ground state wavefunction. Convergence becomes problematic when $J_\perp$ is very small $J_\perp < 0.001$ but
results are reliable for bigger values of the coupling. We use up to 1500 states per block and chain length up to 200 spins.

If we consider the low-lying $S^z=0$ (where $S^z$ is the total spin) state we observe nothing remarkable. The magnetization is uniformly vanishing
and this does not change from the decoupled limit $J_\perp=0$ till $J_\perp$ very large as expected.
We now compute the ground state wavefunction in each sector $S^z=1,2,3$~: we expect that $S^z=1,2$ display only edge modes
while $S^z=3$ should capture a bulk magnon state. We vary the rung coupling in the $S^z=1$ sector in 
figs.(\ref{lad1-003},\ref{lad1-05},\ref{lad1-4}). For small $J_\perp$ we observe very clear edge mode oscillations of the local magnetization
and when the chain is long enough the bulk magnetization is zero (green line in the figures)~: see fig.(\ref{lad1-003}). 
If we increase $J_\perp$ there is coexistence of  edge oscillations and bulk magnetization~: see fig.(\ref{lad1-05}). Finally at large enough
$J_\perp$ all edge phenomena are washed out and we observe a single bump of the magnetization. We interpret this state as a single
magnon state of the infinite rung limit that propagates as a particle in a box as in ref.(\cite{white_numerical_1993}). The important information
is the gradual disappearance of the edge modes without any level crossing.

We expect that the same scheme apply for the lowest-lying $S^z=2$ state. Its behavior is displayed in
figs.(\ref{lad2-003},\ref{lad2-015},\ref{lad2-05},\ref{lad2-4}). here we again observe clear edge modes exponentially localized
at the end of the ladder no bulk magnetization at small $J_\perp$. There is a smooth crossover to a regime that we interpret
as a two-magnon state~: see fig.(\ref{lad2-4}).

The situation should be different for the $S^z=3$ sector because in the decoupled limit $J_\perp=0$ one is forced to excite at least
one magnon in one of the chains. This magnon will be delocalized and its total magnetization should spread all over the chain.
This is exactly what we observe in fig.(\ref{lad3-003},\ref{lad3-05}). The bulk is always magnetized even when $J_\perp\rightarrow0$.
In addition to this magnon contribution there are also edge oscillations. When increasing the coupling the magnon contribution becomes
clearer~: see fig.(\ref{lad3-05}). And finally for very large rung coupling fig.(\ref{lad3-4})
we are left with only rung magnon excitations and no edge modes. This is a gradual phenomenon with no physical discontinuity.

In principle NMR on well-chosen chain-breaking impurities should be able to measure these physical effects.

\section{Conclusion}
Antiferromagnetic spin chains are a fascinating playground for states of magnetic matter that do not fit into the usual
symmetry-breaking views of magnetism. The crafting of molecular magnets has allowed chemistry and physics to join forces
and investigate experimentally the Haldane conjecture that predicts a fundamental difference between integer and half-integer
spin chains. In the integer case the so-called Haldane weakens with the spin value and we have proposed here a conjecture
for its asymptotic behavior for large spin S. Another very special feature of the Haldane gap state is the existence
of long-range correlations that leads to edge modes that have been observed in experiments~\cite{hagiwara_observation_1990}.
These correlations are fragile and can be destroyed even without crossing any phase transition. We have given an explicit
mechanism for the destruction of the hidden order by showing that a S=1 spin ladder interpolates between haldane-style phase
to a simple tensor product phase with no kind of long-range order. The crossover we describe should be accessible
to NMR measurements provided one finds the appropriate molecular magnet.
In this respect the recent studies of BIP-TENO~\cite{katoh_singlet_2000,sakai_magnetization_2003} or other 
magnets~\cite{mennerich_antiferromagnetic_2006} show that it is indeed feasible to study spin-1 ladders and observe 
the interesting crossover between the Haldane phase and the rung singlet phase.

\section{Acknowledgments}
We acknowledge stimulating exchange interactions with M. Verdaguer and J. P. Renard over the years.
We thank Gregoire Misguich for help with DMRG ITensor library~\cite{ITensor}. We thank also 
TOPMAT workshop of PSI2 project funded by the IDEX Paris-Saclay ANR-11-IDEX-0003-02 where part of this work has been performed.
We also thank Ph. Lecheminant for careful reading of our manuscript and K. Totsuka for useful discussions.

\section*{References}

\bibliography{spins_revised}

\end{document}